\documentstyle [prl,aps,epsfig] {revtex}
\begin{document}
\draft
\twocolumn[\hsize\textwidth\columnwidth\hsize\csname
@twocolumnfalse\endcsname
\date \today
\title{Evidence for universal structure in galactic halos}

\author{William H. Kinney and Pierre Sikivie}
\address{Physics Department, University of Florida, Gainesville, FL 32611}

\maketitle

\begin{abstract}
The late infall of dark matter onto a galaxy produces structure (such as 
caustics) in the distribution of dark matter in the halo.  We argue that 
such structure is likely to occur generically on length scales proportional 
to  $l \sim t_0 v_{\rm rot}$, where $t_0$ is the age of the universe and 
$v_{\rm rot}$ is the rotation velocity of the galaxy.  A set of 32 extended 
galactic rotation curves is analyzed.  For each curve, the radial coordinate 
is rescaled according to $r\rightarrow \tilde r \equiv r (v_0 / v_{\rm rot})$,
where we choose $v_0 = 220\,{\rm km/s}$. 
A linear fit to each rescaled rotation curve is subtracted, and the 
residuals are binned and averaged.  The sample shows significant features 
near $\tilde r = 40\,{\rm kpc}$ and $\tilde r = 20\,{\rm kpc}$.  This is 
consistent with the predictions of the self-similar caustic ring model 
of galactic halos.

\end{abstract}

\pacs {PACS numbers: 95.35.+d, 98.35.Gi}
\vskip1pc]

The rotation curves of most spiral galaxies are approximately flat, i.e.
$(v(r) \simeq$ constant) for $r$ much larger than the disk radius \cite{vr}, 
where $v(r)$ 
is the circular velocity of gas at radial coordinate $r$.  The flatness 
of galactic rotation curves implies that galaxies are surrounded by halos 
of dark matter, and that the halo density falls off approximately as 
$\rho_{DM} (r) \propto {1 /r^2}$.  The curves are far from exactly flat, 
however.  They have all sorts of irregularities which may be referred to  
as ``bumps''.  When a bump occurs at $r>>$ disk radius, it indicates 
structure in the dark matter distribution, i.e. a deviation from a perfect 
$\rho_{DM} \propto {1 / r^2}$ law.  We are motivated by the possibility 
that such structures have universality and that a statistical analysis 
of well measured extended rotation curves may reveal that.

In refs. \cite{kg,rh}, 32 extended galactic rotation curves were selected
and analyzed to test the validity of a modification of Newtonian dynamics
(MOND) as an explanation of the ``dark matter problem'', i.e. the discrepancy
between the flatness of galactic rotation curves and the exponential  
fall-off of galactic luminous matter distributions.  We analyze the
same set of rotation curves with a different goal, namely to search for  
structure within dark matter halos.  We adopt the conventional wisdom that 
a rotation curve is a tracer of the galactic mass distribution according to 
the laws of Newtonian dynamics.  However, the selection criteria used 
by the authors of refs. \cite{kg,rh} - that each rotation curve is an 
accurate tracer of the radial force law, and that it extends far beyond 
the edge of the luminous disk - are appropriate from the point of view of 
our analysis.  We do not know of another data set of comparable quality.  
We use the complete data set described there, without any cuts of our own.

One broad argument why structure may be present in galactic halos is that 
the halos form as a result of the infall of the dark matter surrounding the 
galaxy \cite{jg}.  If the dark matter particles are collisionless, they
oscillate back and forth numerous times before they are virialized 
by inhomogeneities in the galactic mass distribution \cite{ip}.  The 
non-virialized flows of dark matter produce structure within the halo, e.g.
the caustics described below.  

If structure is present, one would expect it to occur on length scales set 
by $t~v_{rot}~$ where $t$ is the age of the galaxy and $v_{rot}$ is its 
rotation velocity.  Indeed $v_{rot}$ sets the scale for the velocities with 
which all dark matter particles move about in the halo.  Since all galaxies 
have nearly the same age, we expect that any regularity shared by the 
different galaxies of a set  will reveal itself most readily after all 
lengths have been rescaled according to 
$\ell \rightarrow \ell^\prime = {\ell / v_{rot}}$.

We are motivated in large part by the fact that caustics form in the 
non-virialized flows associated with the late infall of dark matter onto 
a galaxy.  By definition, a caustic is a place in physical space where 
the dark matter density is very large.  In the limit where the infalling 
dark matter has zero velocity dispersion, the dark matter density diverges 
at the caustics.  Two types of caustic form: {\it inner} and {\it outer} 
\cite{sin}.  The outer caustics are 2-spheres surrounding the galaxy.  
One such caustic is located wherever an outflow of dark matter reaches 
its maximum radius before falling back in.  The inner caustics are rings.  
One caustic ring is located wherever the particles with the most angular 
momentum in a given inflow reach their distance of closest approach 
to the galactic center before moving back out.  

Outer caustic spheres and inner caustic rings are examples of the sort of
structure we are looking for.  Caustic rings are the more likely candidate 
for detection in an analysis of galactic rotation curves because they have 
the higher density contrast and because they are located closer to the 
galactic center, where the rotation curves are measured.  
Furthermore, a specific proposal \cite{cr} has been made for the radii 
$a_n$ of caustic rings:
\begin{eqnarray}
\{a_n: n=1,2, ...\} \simeq &&(39,~19.5,~13,~10,~8,...) {\rm kpc} \cr
 &&\times \left({j_{\rm max}\over 0.25}\right) \left({0.7\over h}\right) 
\left({v_{rot} \over 220 {km\over s}} \right)
\label{1}
\end{eqnarray}
where $h$ is the present Hubble rate in units of $100\,{\rm km/(s~Mpc)}$, and 
$j_{\rm max}$ is the maximum of the dimensionless angular momentum 
distribution of the infalling dark matter particles in the self-similar 
infall model of galactic halo formation.  

The infall of dark matter is called self-similar \cite{ss} if it is 
time-independent after all distances are rescaled by a time-dependent 
scale $R(t)$ and all masses are rescaled by the mass $M(t)$ interior to 
$R(t)$.  Usually $R(t)$ is taken to be the turn-around radius, defined 
as follows.  Consider all particles which are about to fall onto the 
galaxy for the first time in their history at time $t$.  Such particles 
are said to be at their `first turn-around'.  In a spherical model, they 
have zero radial velocity then.  Their distance to the galactic center 
is the turn-around radius $R(t)$ at that time.  In the case of zero 
angular momentum and spherical symmetry, the infall is self-similar if 
the initial overdensity profile has the form 
\begin{equation}
{\delta M_i \over M_i} = \left({M_0\over M_i}\right)^\epsilon,
\end{equation} 
where $M_0$ and 
$\epsilon$ are parameters \cite{ss}.  $\epsilon$ must be in the range 
$0 \leq \epsilon \leq 1$.  The rotation curve is flat if 
$0 \leq \epsilon \leq 2/3$ \cite{ss}.  In models of large scale structure
formation, $\epsilon$ is predicted to lie in the range 0.2 to 0.35~
\cite{sty}.  The self-similar infall model was generalized in ref.
\cite{sty} to include the effect of angular momentum.  It was found that, 
in addition to the condition on the initial overdensity profile, 
self-similarity requires the angular momentum distribution $\ell(t)$ to 
have the time-dependence
\begin{equation}
\ell (t) = j \left({R^2(t) \over t}\right),
\end{equation}
where $j$ is a 
dimensionless and time-independent distribution.  The maximum of the 
$j$ distribution is the quantity $j_{\rm max}$ which appears in Eq.~(1).  
Good agreement of the self-similar model with the properties of our 
own galaxy was found \cite{sty} for parameter values $\epsilon = 0.20$
to 0.35, ~$\bar j \simeq 0.2$ and $h \simeq 0.7$ where $\bar j$ is the 
average of the $j$ distribution.  

Eq.~(1) assumes $\epsilon = 0.3$.  For $\epsilon = 0.2$, 
\begin{equation}
a_1 \simeq 36 {\rm kpc}~\left({j_{\rm max}\over 0.25}\right) \left({0.7\over 
h}\right)  
\left({v_{rot}\over 220 km/s}\right),
\end{equation}
 but the ratios $a_n/a_1$ are almost
the same as in the $\epsilon = 0.3$ case.  The ratios happen to be 
close to $a_n/a_1 = 1/n$ over the range ($0.2 \leq \epsilon \leq 0.35$)
of interest.  

Eq.~(1) predicts the caustic ring radii of a galaxy in terms of its first 
ring radius $a_1$.  If the caustic rings lie close to the galactic plane 
they cause bumps in the rotation curve at the caustic ring radii.  
As a possible example of this effect, consider \cite{cr} the rotation 
curve of NGC3198, one of the best measured and a member of the set selected 
in refs. \cite{kg,rh}.  It has three faint bumps at radii: 28, 13.5 and 9 kpc, 
assuming $h=0.75$.  The ratios happen to be consistent with Eq.(1) assuming 
the bumps are caused by the first three ($n=1,2,3$) ring caustics of NGC3198.  
Moreover, since $v_{\rm rot} = 150$ km/s, $j_{\rm max}$ is determined in 
terms of $\epsilon$.  For $\epsilon = 0.3,~j_{\rm max} = 0.28$.  Note that 
the uncertainty in $h$ is a systematic effect that can be corrected for when 
determining $j_{\rm max}$ because the bump radii scale like $1/h^\prime$ 
where $h^\prime$ is the Hubble rate assumed by the observer in constructing 
the rotation curve, and the caustic ring radii scale as $1/h$.  Rises in 
the inner rotation curve of the Milky Way were also interpreted \cite{cr} 
as due to caustics $n =6,7,8,9,10,11,12$ and 13.  This determined 
the value of $j_{\rm max}$ of our own galaxy to be 0.263 for $\epsilon = 0.3$, 
the value we assume henceforth.  The first five caustic ring radii in our 
galaxy are then predicted to be: 41,~20,~13.3,~10,~8~kpc. 
    
According to the self-similar caustic ring model, each galaxy has its own
value of $j_{\rm max}$.  Over the set of 32 galaxies selected in refs. 
\cite{kg,rh}, $j_{\rm max}$ has some unknown distribution.  However, the 
fact that the values of $j_{\rm max}$ of NGC3198 and of the Milky Way happen 
to be close to one another, within 7\%, suggests that the $j_{\rm max}$ 
distribution may be peaked near a value of 0.27~.  Our strategy is to rescale 
each rotation curve according to 
\begin{equation}
r \rightarrow \tilde r  = r\left({220\,{\rm km/s} \over v_{rot}}\right)
\end{equation}
and to add them together in a way made precise below.  Since Eq.~(1) predicts 
the $n$th caustic radius $a_n$ to be distributed like $j_{\rm max}$ for all n, 
and it fixes the ratios $a_n/a_1 \simeq 1/n$, the sum of rotation curves 
should show the $j_{\rm max}$ distribution, once for $n=1$, then at about 
half the $n=1$ radii for $n=2$, then at about 1/3 the $n=1$ radii for $n=3$, 
and so on.  If the $j_{\rm max}$ distribution is broad, the sum of rotation 
curves is unlikely to show any feature.  However, if it is peaked, then the 
sum should show a peak for $n=1$ at some radius, then again at 1/2 that 
radius for $n=2$, at 1/3 the radius for $n=3$, and so on.  If the 
$j_{\rm max}$ distribution is peaked at 0.263 (the value for the Milky 
Way), the peaks in the sum of rotation curves should appear at 41 kpc, 
20 kpc, 13.3 kpc, and so forth.

We search for overdensities in the halo dark matter distribution as 
manifested by upward fluctuations in velocity relative to the underlying 
flat rotation curve of the galaxy.  Unfortunately, rotation curves are 
never exactly flat.  They often rise or fall systematically with increasing 
radius, and all show a sharp drop in rotation velocity near the galactic 
center.  Furthermore, individual sources (i.e. gas) in a galaxy have 
peculiar velocities independent of any caustic structure in the galaxy. 
We require a definite procedure for subtracting out the background 
rotation of the galaxy, and for quantifying the ``noise'' due to the 
peculiar velocities of sources.  The procedure we adopt is as follows.  
Each rotation curve is expressed as a set of radii $r_i$ in kpc
and rotation velocities $v_i$ in ${\rm km/sec}$.  Each ($r_i,~v_i$) 
is given by a data point in the rotation curves published in refs. 
\cite{kg,rh}.\footnote{These curves are heavily processed and do not 
represent the velocities of individual sources. See, for example, 
Ref.\cite{martimbeau94} and references therein for a discussion of 
the observational details relating to one galaxy in our sample, IC2574.
In the case of the galaxy M33, whose published rotation curve 
\cite{kg,rh} is so densely sampled that the data points overlap with 
one another, we choose to sample the published curve at intervals of 
$0.25\,{\rm kpc}$, a sufficient resolution for our purpose here.}  
Now, in the case  of a flat rotation curve, $v_{\rm rot}$ is just the 
average velocity of the outer rotation curve, which we denote $\bar v$.  
In the case of a rising or falling rotation curve, the ``overall'' rotation 
velocity $v_{\rm rot}$ is less precisely defined. We adopt the prescription 
$v_{\rm rot} \equiv \bar v$ in all cases.  Thus the rescaled radii 
$\tilde r_i$ are
\begin{equation}
\tilde r_i \equiv r_i  \left({220\,{\rm km/s} \over \bar v}\right),
\end{equation}
where $\bar v$ is calculated according to the following procedure.  We are 
interested in the outer, approximately flat portion of the rotation curve, so
we remove the inner points by specifying a cut: all points with {\it rescaled} 
radii $\tilde r_i <10\,{\rm kpc}$ are thrown out. The remaining points in the 
rotation curve are then fitted to a line, and $\bar v$ is the corresponding 
average.  A subtlety arises because the average velocity $\bar v$ is 
calculated after the cut is applied, but the cutoff is defined in rescaled 
coordinates and therefore depends on $\bar v$.  The cutoff procedure is in 
fact performed iteratively: we start with a guess for $\bar v$, calculate 
the cutoff radius, recalculate $\bar v$ based on the cutoff radius, and 
so on until both quantities converge (usually after one or two iterations). 

Fitting the outer rotation curve to a line may seem an arbitrary choice. In 
practice, it works quite well: deviations from a linear fit are typically 
less than $10\,{\rm km/s}$ in galaxies with a typical rotation velocity of 
$200\,{\rm km/s}$.  However, to verify the robustness of our conclusions, 
we also performed the analysis with quadratic polynomial fits to the 
rotation curves, with no substantial change in the results.  (Of course, a 
fit to a high enough order polynomial will remove any features present in 
the rotation curve!).

\begin{figure}
\centerline {\epsfysize=3.8in \epsfbox{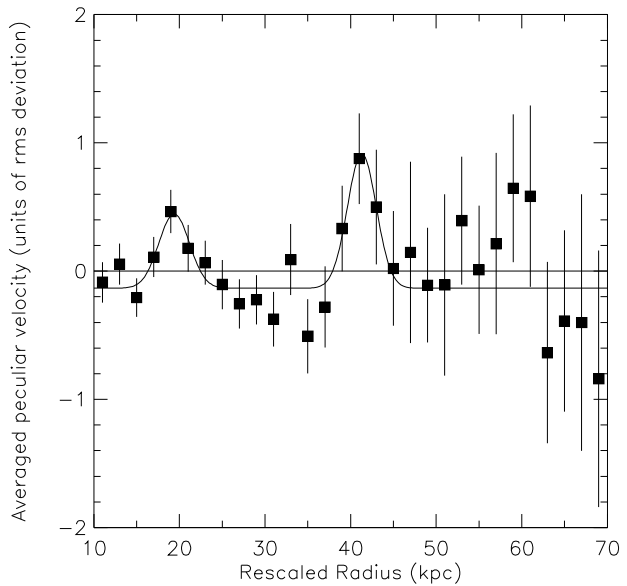} }
\caption{Binned data for 32 galaxy sample, with peaks fit to Gaussians}
\label{figbinneddata}
\end{figure}

Once the linear fit is calculated, this background rotation is subtracted 
off the rotation curve, leaving a set of peculiar velocities $\delta v_i$. 
It is then straightforward to calculate an rms ``noise'' 
$\sqrt{\left\langle \delta v~^2 \right\rangle}$ for each galaxy,
\begin{equation}
\sqrt{\left\langle \delta v~^2 \right\rangle} \equiv 
\sqrt{{1 \over N - N_{\rm fit}} \sum_{i} \left(\delta v_i\right)^2},
\end{equation}
where $N$ is the number of points and $N_{\rm fit}$ is the number of degrees 
of freedom in the fit ($N_{\rm fit} = 2$ for a fit to a line). We adopt 
$\sqrt{\left\langle \delta v~^2 \right\rangle}$ as the size of the error bar 
on the $\delta v_i$ for a given galaxy.  This error is a measure of the 
intrinsic peculiar velocities of the sources, which we assume to be  random. 
It is considerably more conservative than the quoted observational errors on 
the points in the rotation curve.  Finally, the peculiar velocities are 
expressed in dimensionless units 
$\delta \tilde v_i \equiv v_i / \sqrt{\left\langle \delta v~^2 \right\rangle}$, 
and the sample of galaxies is averaged into radial bins:
\begin{equation}
b_i \equiv {1 \over N_i} \sum_{j=1}^{N_i} \delta \tilde v_j,
\end{equation}
where $N_i$ is the number of data points in the $i$-th bin. The assigned error 
on each $b_i$ is then simply $1 /\sqrt{N_i}$. Figure \ref{figbinneddata} 
shows the complete set of 32 galaxies averaged into $2\,{\rm kpc}$ bins. 

There are two features evident at roughly $20$ and $40\,{\rm kpc}$. A fit to 
two Gaussians (with amplitude, width and mean left as free parameters) 
plus a constant indicates features at $19.4 \pm 0.7\,{\rm kpc}$ and 
$41.3 \pm 0.8\,{\rm kpc}$, with overall significance of $2.4\sigma$ and 
$2.6\sigma$, respectively.  Figure \ref{figbinneddata} also shows the fitted 
curve.  When the same fit is applied to the same data in 1 kpc bins, the 
significance of the two peaks is $2.6 \sigma$  and $3.0 \sigma$ respectively.
The locations of the features agrees with the predictions of the 
self-similar caustic ring model with the $j_{\rm max}$ distribution peaked 
at 0.27.  The use of Gaussians to fit the peaks in the combined rotation 
curve was an arbitrary choice in the absence of information on the 
$j_{\rm max}$ distribution.

We have shown that there is evidence for universal structure in the dark 
halos of spiral galaxies, in that the rotation velocity has a statistically 
significant tendency to fluctuate upward at rescaled radii $\tilde r \sim 20$ 
and $40\,{\rm kpc}$.  While the significance of these peaks in the data set 
analyzed is not overwhelming, roughly $2.5\sigma$ for each peak when 
considered with appropriately conservative error bars, our analysis provides 
a tantalizing indication that the structure is real and that it is within 
observational reach of a more detailed survey of galactic rotation curves.  

It is particularly striking that the positions of the peaks coincide 
with the radii of the $n=1$ and $n=2$ caustic rings in the self-similar
infall model for $j_{\rm max} \sim 0.27$.  This strongly suggests that the 
$j_{max}$ distribution is indeed peaked near the value of 0.27, as the 
earlier estimates of the $j_{max}$ values of NGC3198 (0.28) and the Milky 
Way (0.265) had suggested.

The authors would like to thank Francis Halzen and John Yelton for 
invaluable conversations and suggestions.   This work is supported 
in part by U.S. DOE grant DEFG05-86ER-40272.

\end{document}